\documentclass[aps,prx,final,10pt,twocolumn,superscriptaddress]{revtex4-2}
\usepackage{amsmath,amssymb,bm}
\usepackage{graphicx,color}
\usepackage[squaren]{SIunits}
\usepackage[english]{babel}
\usepackage{amstext}
\usepackage{amsthm}
\usepackage{latexsym}
\usepackage{array}
\usepackage{color}
\usepackage{float}
\usepackage{microtype}
 
\usepackage{multirow}
\usepackage{dcolumn}
\usepackage{soul}
\usepackage{lettrine}
\usepackage{type1cm} 
\usepackage{natbib}
\usepackage{longtable}
\definecolor{url}{RGB}{0,20,160}
\usepackage[colorlinks=true,linkcolor=blue,citecolor=blue,urlcolor=url]{hyperref}
\usepackage[usenames,dvipsnames,svgnaes,table]{xcolor}
\makeatletter
\def\frutiger{cmss10 }
\def\frutigerbold{cmssbx10 }
=\frutigerbold at 15pt
=\frutiger at 10pt
=\frutigerbold at 12pt
=\frutigerbold at10pt
=\frutigerbold at 8pt
=\frutiger at 8pt
=\frutigerbold at 31pt
\def\@caption@tabnum@sep{\figtextfont{{ }{\bf\textbar}{ }}}%
\def\fnum@table{{\bf\tablename~\thetable}}

\def\@caption@fignum@sep{\figtextfont{{ }{\bf\textbar}{ }}}%
\def\fnum@figure{{\bf\figurename~\thefigure}}
\renewenvironment{figure}{\@float{figure}\def\textbf##1{{\fignumfont ##1}}\def\bf{\fignumfont}}{\end@float}
\def\@startsection#1#2#3#4#5#6{%
	\if@noskipsec\leavevmode\fi
	\par\@tempskipa #4\relax
	\@afterindenttrue
	\ifdim\@tempskipa <\z@
	\@tempskipa -\@tempskipa \@afterindentfalse
	\fi\if@nobreak\everypar{}%
	\else\addpenalty\@secpenalty\addvspace\@tempskipa\fi
	\@ifstar{\@ssect{#3}{#4}{#5}{#6}}{\@dblarg{\@sect{#1}{#2}{#3}{#4}{#5}{#6}}}}
\def\@sect#1#2#3#4#5#6[#7]#8{%
	\ifnum #2>0
	\let\@svsec\@empty
	\else\refstepcounter{#1}\protected@edef\@svsec{\@seccntformat{#1}\relax}\fi
	\@tempskipa #5\relax
	\ifdim\@tempskipa>\z@
	\begingroup#6{\@hangfrom{\hskip #3\relax\@svsec}%
		\interlinepenalty \@M #8\@@par}\endgroup
	\csname #1mark\endcsname{#7}%
	\addcontentsline{toc}{#1}{%
		\ifnum #2>\c@secnumdepth\else
		\protect\numberline{\csname the#1\endcsname}\fi #7}%
	\else\def\@svsechd{#6{\hskip #3\relax
			\@svsec #8\ifnum#2=2.\fi}%
		\csname #1mark\endcsname{#7}%
		\addcontentsline{toc}{#1}{%
			\ifnum #2>\c@secnumdepth \else
			\protect\numberline{\csname the#1\endcsname}\fi #7}}%
	\fi\@xsect{#5}}

\renewcommand\section{\@startsection {section}{1}{\z@}%
	{-10pt \@plus -1ex \@minus -.2ex}{.5ex }{\normalfont\Large\bfseries\sectionfont}}
\renewcommand\subsection{\@startsection{subsection}{2}{\z@}%
	{10pt\@plus 1ex \@minus .2ex}{-0.5ex \@plus .2ex}{\normalfont\large\bfseries\subsectionfont}}
\def\frontmatter@title@format{\titlefont\centering}%
\def\frontmatter@title@below{\addvspace{-5pt}}%

\renewcommand\NAT@biblabelnum[1]{#1.}
\renewcommand\NAT@citesuper[3]{\ifNAT@swa
	\unskip\hspace{1\p@}\textsuperscript{(#1)}%
	\if\relax#3\relax\else\ (#3)\fi\else (#1)\fi\endgroup}

\newcommand*\bib@heading{%
	\section{\refname}
	\fontsize{8}{10}\selectfont
}
\newcommand*\@openbib@code{%
	\advance\leftmargin\bibindent
	\itemindent -\bibindent
	\listparindent \itemindent
	\parsep \z@
}%
\newdimen\bibindent
\usepackage[usenames,dvipsnames,svgnaes,table]{xcolor}
\definecolor{col1}{rgb}{0.0, 0.30, 1.0}
\definecolor{col2}{rgb}{0.9, 0.0, 0.30}

\usepackage[draft]{changes} 
\usepackage[normalem]{ulem}
\definechangesauthor[name={yi}, color=red]{Yi}

\makeatletter

\newcommand{\Rmnum}[1]{\expandafter\@slowromancap\romannumeral #1@}
\newcommand{\revise}[1]{{\textcolor{red}}}

\begin{document}
\title{Realizing Intrinsically Glass-like Thermal Transport via Weakening the Ag-Ag Bonds in Ag$_{6}$ Octahedra}

\author{Xingchen Shen}
\email{xingchen.shen@ensicaen.fr}
\affiliation{CRISMAT, CNRS,ENSICAEN,14000,Caen,France}
\altaffiliation{Contributed equally to this work}

\author{Zhonghao Xia}
\affiliation{Key Laboratory of Advanced Materials and Devices for Post-Moore Chips, Ministry of Education, University of Science and Technology Beijing, Beijing 100083, China}
\affiliation{School of Mathematics and Physics, University of Science and Technology Beijing, Beijing 100083, China}
\affiliation{Contributed equally to this work}

\author{Jun Zhou}
\affiliation{Basic Experimental Center for Natural Science, University of Science and Technology Beijing, Beijing 100083, China}

\author{Yuling Huang}
\affiliation{Department of Mechanical and Energy Engineering, Southern University of Science and Technology (SUSTech), Shenzhen, 518055 China}

\author{Yali Yang}
\affiliation{School of Mathematics and Physics, University of Science and Technology Beijing, Beijing 100083, China}

\author{Jiangang He}
\email{jghe2021@ustb.edu.cn}
\affiliation{Key Laboratory of Advanced Materials and Devices for Post-Moore Chips, Ministry of Education, University of Science and Technology Beijing, Beijing 100083, China}
\affiliation{School of Mathematics and Physics, University of Science and Technology Beijing, Beijing 100083, China}

\author{Yi Xia}
\email{yixia@pdx.edu}
\affiliation{Department of Mechanical \& Materials Engineering, Portland State University, Portland, OR 97201, USA}

\begin{abstract}
\noindent Crystals exhibiting glass-like and low lattice thermal conductivity ($\kappa_{\rm L}$) are not only scientifically intriguing but also practically valuable in various applications, including thermal barrier coatings, thermoelectric energy conversion, and thermal management. However, such unusual $\kappa_{\rm L}$ are typically observed only in compounds containing heavy elements, with large unit cells, or at high temperatures, primarily due to significant anharmonicity. In this study, we utilize chemical bonding principles to weaken the Ag-Ag bonds within the Ag$_6$ octahedron by introducing a ligand in the bridge position. Additionally, the weak Ag-chalcogen bonds, arising from fully filled $p$-$d$ antibonding orbitals, provide an avenue to further enhance lattice anharmonicity. We propose the incorporation of a chalcogen anion as a bridge ligand to promote phonon rattling in Ag$_6$-octahedron-based compounds. Guided by this design strategy, we theoretically identified five Ag$_6$ octahedron-based compounds, $A$Ag$_3X_2$ ($A$ = Li, Na, and K; $X$ = S and Se), which are characterized by low average atomic masses and exhibit exceptionally strong four-phonon scattering. Consequently, these compounds demonstrate ultralow thermal conductivities (0.3 $\sim$ 0.6 Wm$^{-1}$K$^{-1}$) with minimal temperature dependence (T$^{-0.1}$) across a wide temperature range. Experimental validation confirmed that the $\kappa_{\rm L}$ of NaAg$_3$S$_2$ is 0.45 Wm$^{-1}$K$^{-1}$ within the temperature range of 200 to 550 K. Our results clearly demonstrate that weak chemical bonding plays a crucial role in designing compounds with glass-like $\kappa_{\rm L}$, highlighting the effectiveness of chemical bonding engineering in achieving desired thermal transport properties.

\end{abstract}
	
	\maketitle
	\section{INTRODUCTION}
	Lattice thermal conductivity is a fundamental property of solids, and crystalline dielectric materials exhibiting extreme $\kappa_{\rm L}$ values are essential for various technological applications, including thermoelectrics, thermal management, and thermal barrier coatings~\cite{padture2002thermal,science5895,qian2021phonon}. Within the framework of the phonon gas model~\cite{peierls1929kinetischen}, heat carriers are identified as propagating phonons, while heat resistance primarily arises from intrinsic phonon-phonon scattering. At temperatures exceeding the Debye temperature, Umklapp phonon-phonon interactions generally dominate the scattering processes~\cite{peierls1996quantum}. As a result, $\kappa_{\rm L}$ typically decreases rapidly with increasing temperature, often following the relationship $\kappa_{\rm L}$ $\propto$ T$^{-1}$~\cite{tritt2005thermal}. However, certain materials deviate from this T$^{-1}$ behavior, exhibiting temperature-independent $\kappa_{\rm L}$ under specific conditions, which reflects glass-like thermal conductivity characteristics. While low and temperature-independent $\kappa_{\rm L}$ is a defining feature of amorphous materials~\cite{zhou2020thermal}, it is very rarely observed in certain crystalline solids, primarily due to the lack of periodicity in amorphous phases~\cite{qian2021phonon,simoncelli2019unified,beekman2017inorganic}.

	It is noteworthy that glass-like thermal conductivities have occasionally been observed in crystalline solids, particularly when the structures are complex, or when impurities and disordered atoms are present, or at sufficiently high temperatures~\cite{beekman2017inorganic,annurev.pc.39}. Compounds exemplifying partial atomic disorder include zinc-disordered $\beta$-Zn$_4$Sb$_3$~\cite{snyder2004disordered}, liquid-like Cu$_{2-x}$Se~\cite{liu2012copper}, and Cu- and Ag-based argyrodites~\cite{https://doi.org/10.1002/advs.202400258,weldert2014thermoelectric}. In the case of $\beta$-Zn$_4$Sb$_3$, Zn interstitials exhibit significant mobilities, diffusing throughout the structure and leading to glass-like thermal conductivity~\cite{snyder2004disordered,10.1063/1.3599483}. In Cu$_{2-x}$Se, the highly disordered Cu atoms surrounding the Se sublattice are superionic and possess liquid-like mobility, contributing to the observed glass-like thermal conductivity~\cite{liu2012copper}. The ultralow thermal transport in crystalline argyrodite Cu$_7$PS$_6$ is attributed to the softening of phonon modes, arising from disorder in the molten copper sublattice~\cite{https://doi.org/10.1002/advs.202400258,weldert2014thermoelectric}. Since anharmonicity typically increases with temperature, the glass-like thermal conductivity is expected at high enough temperature, as long as the lattice is not melted. For example, La$_2$Zr$_2$O$_7$ exhibits crystal-like thermal conductivity at room temperature but glass-like thermal conductivity at the temperature higher than 900 K~\cite{yang2016effective}.
	
	To date, only a limited number of crystalline solids devoid of impurities and disorder exhibit glass-like $\kappa_{\rm L}$. Most of these materials are layered compounds or systems characterized by large and complex unit cells, such as clathrates and skutterudites~\cite{10.1063/1.121747,B706808E}. In these cage-like structures, the relative sizes of the guest atoms and the cages are crucial factors influencing whether a specific composition manifests glass-like or crystal-like $\kappa_{\rm L}$. This relationship is linked to the displacement amplitudes of the guest atoms, which in turn affects the strength of anharmonic phonon-phonon scattering. Notably, layered halide perovskites, such as Cs$_3$Bi$_2$Br$_9$ and Cs$_3$Bi$_2$I$_6$Cl$_3$, have been found to exhibit extremely low glass-like thermal conductivities~\cite{li2024phonon,acharyya2022glassy}, attributable to their heavy atomic mass, low acoustic phonon frequencies and sound velocities.

	Interestingly, Cu$_{12}$Sb$_{4}$S$_{13}$, which exhibits neither a cage-like nor a layered structure, demonstrates an exceptionally low $\kappa_{\rm L}$ with glass-like behavior ranging from 200 to 600 K~\cite{lu2013high,suekuni2018retreat}. A previous theoretical study revealed that the temperature-induced hardening of low-lying optical modes effectively mitigates the scattering of heat-carrying acoustic modes, which is otherwise exacerbated by the rattling of Cu cations within the CuS$_3$ triangle. This mitigation occurs through the reduction of available phase space for three-phonon processes, thereby counterbalancing the conventional $\propto$ T increase in scattering associated with phonon population and resulting in a nearly temperature-independent $\kappa_{\rm L}$~\cite{PhysRevLett.125.085901}. The rattling of Cu within the CuS$_3$ triangle can be attributed to the $p$-$d$ antibonding interactions between the coinage metal cation and the chalcogenide anion~\cite{https://doi.org/10.1002/adfm.202108532}, which leads to chemical bond softening and atomic rattling. Similarly, the observed low acoustic phonon frequencies and sound velocities in Cs$_3$Bi$_2$Br$_9$ and Cs$_3$Bi$_2$I$_6$Cl$_3$ can also be ascribed to the weak bonding interactions between Bi and the halide cations, alongside their substantial atomic masses. These findings suggest that weak chemical bonds may offer a promising avenue for the design and discovery of compounds characterized by glass-like $\kappa_{\rm L}$.

    In this study, we propose a strategy for designing crystalline solids with glass-like $\kappa_{\rm L}$ based on chemical bonding principles. Guided by this design strategy, we identified five Ag$_6$ octahedron-based semiconductors that exhibit exceptionally low $\kappa_{\rm L}$ values and glass-like thermal conductivities within the temperature range of 200 to 550 K, using first-principles calculations. We experimentally validated our predictions for NaAg$_3$S$_2$. Our results, derived from state-of-the-art calculations and experimental investigations, demonstrate that NaAg$_3$S$_2$ possesses a $\kappa_{\rm L}$ of 0.45 W m$^{-1}$ K$^{-1}$ across the temperature range of 200 $\sim$ 550 K. This behavior is attributed to the interplay between flat-band phonon hardening and increased three- and four-phonon scattering rates with rising temperature, which arise from the antibonding interactions between Ag and S, as well as the weak chemical bonding between Ag atoms in Ag$_6$ octahedron. Furthermore, our calculations indicate that KAg$_3$S$_2$ exhibits a similar thermal conductivity behavior, with an even lower $\kappa_{\rm L}$ of 0.23 W m$^{-1}$ K$^{-1}$. The rattling motion of Ag atoms disrupts the conventional phonon-gas model, underscoring the significant role that wave-like (phonon tunneling) thermal conductivity plays in shaping $\kappa_{\rm L}$.

	\section{METHODS}
	\noindent \textbf{Computational methods.} The DFT calculations of lattice dynamics conducted in this work were carried out using the Vienna \textit{ab initio} Simulation Package (VASP)~\cite{vasp1,vasp2}. We utilized the projector augmented wave (PAW) method~\cite{PAW1,PAW2} with a plane wave basis set, employing a cutoff energy of 520 eV. The calculations were performed using the PBEsol version~\cite{PBEsol} of the generalized gradient approximation (GGA) exchange-correlation functional. To sample the Brillouin zone, we employed $\Gamma$-centered $k$-point grids with a density exceeding 8000 $k$-points per reciprocal atom (KPPRA). All structures were fully relaxed, with the forces on each atom constrained to be smaller than 0.01 eV/\AA, and the total energy convergence criterion was set to 10$^{-8}$ eV. To elucidate the nature of chemical bonding, we performed Crystal Orbital Hamilton Population (COHP) analysis for various pairs of nearest neighbor atoms using the LOBSTER code~\cite{nelson2020lobster}. The bulk modulus ($B$), shear moduli ($G$), and average speed of sound ($\nu_a$) were calculated from the elastic constants obtained through finite differences methods.
	
    The second-order force constants (2ndFC) were calculated using the finite displacement method as implemented in the Phonopy code~\cite{TOGO20151}. A \(2 \times 2 \times 2\) supercell and a \(2 \times 2 \times 2\) $k$-point mesh were utilized, with a displacement of 0.01 \AA. The projection operators were evaluated in the reciprocal space (LREAL = FALSE) to enhance the accuracy of the force calculations. Random configurations were generated from 20,000 steps of \textit{ab initio} molecular dynamics (AIMD) simulations based on machine learning force fields (MLFF),\cite{PhysRevB.100.014105} using a time step of 1 fs and a temperature of 300 K, from which 20 structures were selected. Self-consistent calculations were subsequently performed to obtain displacement and force datasets. Additionally, compressive sensing lattice dynamics (CSLD)~\cite{RN38} were employed to extract third- and fourth-order force constants. The cutoff distances for the third- and fourth-order interatomic force constants (IFCs) were limited to the sixth-nearest and next-nearest neighbors, respectively.

	The anharmonically renormalized phonon frequencies at finite temperatures were calculated using self-consistent phonon (SCPH) theory~\cite{doi:10.1080/14786440408520575,PhysRevLett.17.753,werthamer1970self}. The thermal transport properties were obtained by solving the Peierls-Boltzmann transport equation through an iterative scheme implemented in the FourPhonon package~\cite{HAN2022108179}. To ensure convergence, uniform \(18 \times 18 \times 18\) and \(14 \times 14 \times 14\) \(q\)-point meshes were employed to calculate the contributions from SCPH three-phonon scattering \((\kappa_{\rm 3ph}^{\rm P})\) and four-phonon scattering \((\kappa_{\rm 3,4ph}^{\rm P})\), respectively. Additionally, the calculation of the four-phonon scattering processes was accelerated using the sampling method~\cite{guo2024sampling}. The off-diagonal contributions from three-phonon scattering \((\kappa_{\rm 3ph}^{\rm C})\) and four-phonon scattering \((\kappa_{\rm 3,4ph}^{\rm C})\) were computed following the formalism developed by Simoncelli et al.~\cite{simoncelli2019unified}.

	\vspace{0.2 cm}
	\noindent \textbf{Experimental methods.} Polycrystalline samples of NaAg$_3$S$_2$ were synthesized using a melting and quenching technique. Stoichiometric amounts of sodium sulfide (Na$_2$S, powder, 99.00\%), silver (Ag, shot, 1-3 mm, 99.99\%), and sulfur (S, pieces, 99.99\%) were accurately weighed and placed into a graphite tube. This tube was subsequently loaded into a silica tube and sealed under a vacuum of approximately 10$^{-4}$ Pa. The sealed tube was heated to 1073 K over a period of 1000 minutes and maintained at this temperature for 1440 minutes, followed by rapid quenching in cold water. The resulting ingots were ground into fine powders and further densified by hot-pressing (HP) at 753 K for 15 minutes under a pressure of 30 MPa. The hot-pressed pellet achieved a density of 96\% of the theoretical density.

    Powder X-ray diffraction (PXRD) of the hot-pressed NaAg$_3$S$_2$ sample at room temperature was conducted using a Bruker D8 diffractometer, employing Cu K$_{\alpha1}$ radiation ($\lambda$ = 1.5406 \AA). Rietveld refinement was performed using the JANA20201 crystallographic computing system, based on the structural model reported by Huster et al.~\cite{petvrivcek2023jana2020,huster1993naag3s2}. As illustrated in Figure 1(c), the Rietveld refinement of the PXRD pattern confirmed that our sample adopts the $Fd\bar{3}m$ (No. 227) crystal structure symmetry with a lattice parameter $a$ = 12.3602(2) \AA. The refined results, including atomic positions, isotropic atomic displacement parameters ($\mathrm{U_{iso}}$), atomic occupancy, and other related information, are tabulated in Table \textcolor{red}{S3}.

    The thermal conductivity ($\kappa$) was calculated using the formula $\kappa$ = $\rho$C$_pd$. The thermal diffusivity ($d$) of the polycrystalline NaAg$_3$S$_2$ sample was assessed using a Netzsch LFA 457 laser flash system under a nitrogen atmosphere, spanning temperatures from 300 to 550 K. Low-temperature measurements of $\kappa$ and heat capacity (C$_p$) below 300 K were obtained using a Quantum Design Physical Property Measurement System (PPMS). The theoretical estimation of C$_p$ above 300 K for NaAg$_3$S$_2$ is 0.36 Jg$^{-1}$K$^{-1}$ according to the Dulong-Petit approximation, and the density ($\rho$) was determined using the Archimedes method. The C$_p$/T versus T$^2$ plot (see Figure~\textcolor{red}{S3}) was fitted using the Debye-Einstein model, with the fitting parameters detailed in Table \textcolor{red}{S4}. Further discussion on this model and the fitting process can be found in the referenced work~\cite{https://doi.org/10.1002/advs.202400258}.

	\begin{figure}[tph!]
	\includegraphics[width=0.9\linewidth]{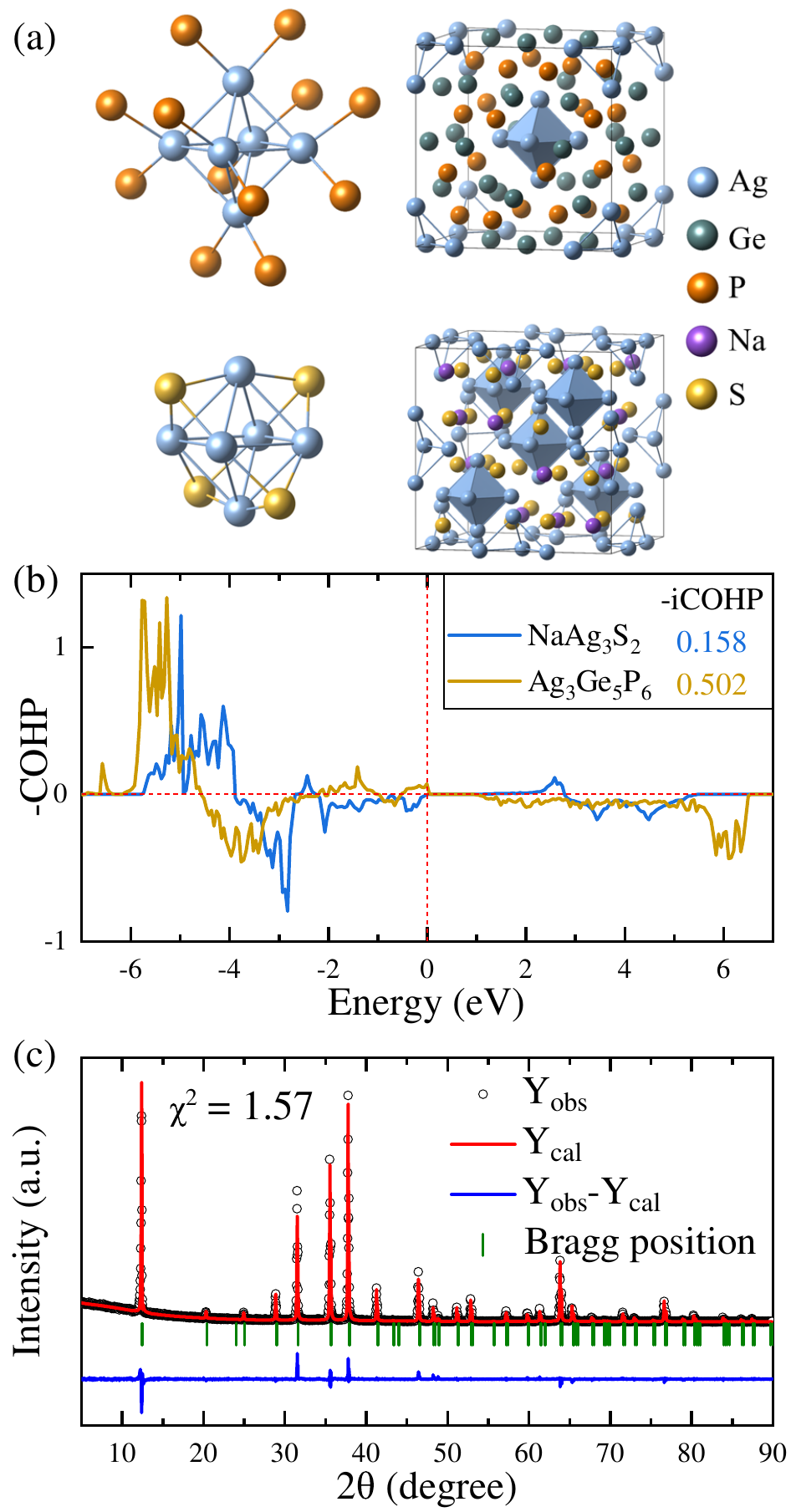}
	\caption{(a) The left panel is Ag$_6$ octahedral and S or P atom bonding modes, and the right panel is the crystal structure of NaAg$_3$S$_2$ and Ag$_3$Ge$_5$P$_6$, wherein an Ag$_6$ octahedral cluster is highlighted. (b) -COHP and -iCOHP of the Ag-Ag bond in the Ag$_6$ octahedron of NaAg$_3$S$_2$ ang Ag$_3$Ge$_5$P$_6$. (c) The refined PXRD of the hot-pressed powder NaAg$_3$S$_2$ sample at 300 K.}
	\label{crystalstructure}
    \end{figure}

	\section{RESULTS AND DISCUSSION}
	\noindent \textbf{Materials design strategies and new materials discovery.} Previous studies indicate that Ag$_3$Ge$_{5}$P$_{6}$ exhibits low $\kappa_{\rm L}$ of 1.7 Wm$^{-1}$K$^{-1}$ at 300 K, with a temperature dependence of 1/T$^{0.59}$ ranging from 200 to 600 K~\cite{C8TA08448C,doi:10.1021/acs.chemmater.7b02474}. This anomalous behavior in lattice heat transport has been primarily attributed to the presence of flat phonon bands, predominantly involving Ag$_6$ octahedra~\cite{xia2019impact}. This is further evidenced by the large mean square atomic displacements (MSD) of Ag atoms and the hardening of the low-lying flat phonon bands as the temperature increases~\cite{xia2019impact,doi:10.1021/acs.chemmater.7b02474}. It has been established that the weak Ag-Ag bond plays a significant role in scattering heat-carrying acoustic phonons, thereby contributing to the unusual heat transfer behavior~\cite{xia2019impact}. The weakness of the Ag-Ag bond arises from the hybridization of the $d$ orbitals with the $s$ and $p$ orbitals of Ag$^{+}$ cation~\cite{doi:10.1021/ic00285a022}. This observation prompts crucial follow-up questions: Is it possible to further weaken the Ag-Ag bond? Could such modifications potentially result in a lower $\kappa_{\rm L}$ and a flatter temperature dependence of the $\kappa_{\rm L}$ - T curve?

    In 1990, Cui and Kertesz demonstrated that the presence of a bridge ligand $X$ between two metal $M$ ions with a $d^{10}$ electron configuration significantly reduces the $M$-$M$ bonding interaction compared to a $M$-$M$ bond without the bridge ligand $X$~\cite{doi:10.1021/ic00339a009}. This reduction is attributed to the filling of the antibonding states formed by the $d$ orbitals of $M$ and the $p$ orbitals of $X$. Based on this finding, it is anticipated that introducing a bridge ligand $X$ in the Ag-Ag bond of an Ag$_6$ octahedron could weaken the Ag-Ag interactions further, potentially leading to enhanced anharmonicity and a lower $\kappa_{\rm L}$. Guided by this design principle, we conducted a search for suitable compounds within the silver chalcogen chemical space using the Inorganic Crystal Structure Database (ICSD)~\cite{icsd1}. The objective was to identify materials that exhibit low $\kappa_{\rm L}$ with reduced temperature dependence.

	\begin{figure}[tph!]
	\includegraphics[width=1.0\linewidth]{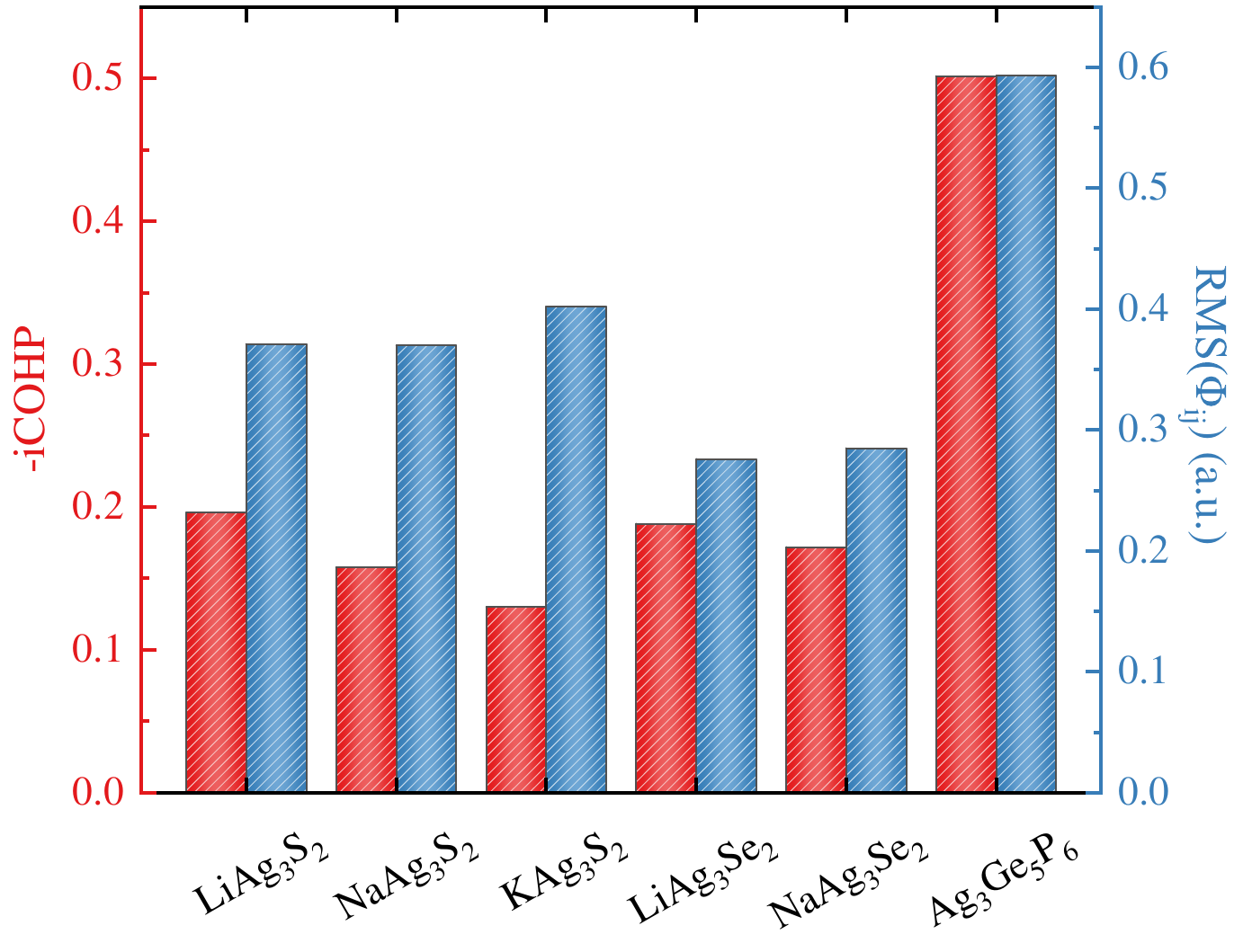}
	\caption{The Ag-Ag bond -iCOHP and $\Phi_{ij}$ of Ag$_3$Ge$_5$P$_6$ and NaAg$_3$S$_2$-type compounds.}
	\label{2ndFC}
    \end{figure}

    We identified two experimentally synthesized and isostructural compounds, NaAg$_3$S$_2$ and KAg$_3$S$_2$, that feature Ag$_6$ octahedra with S$^{2-}$ anions equally bonding to three Ag$^{+}$ cations on one face of the octahedron (see Figure~\ref{crystalstructure}). NaAg$_3$S$_2$ was first synthesized by Huster et al.~\cite{huster1993naag3s2}, while KAg$_3$S$_2$ was introduced by Wood et al.~\cite{C39930000235}. Both compounds crystallize in the cubic system with the space group $Fd\bar{3}m$ (No. 227). The Na$^{+}$/K$^{+}$, Ag$^{+}$, and S$^{2-}$ ions occupy the 16$c$, 48$f$, and 32$e$ Wyckoff positions, respectively, with site symmetries of $\bar{3}m$, $2mm$, and $3m$. In these structures, Na$^{+}$/K$^{+}$ ions are located at the center of the octahedron formed by S$^{2-}$ anions, while six Ag$^{+}$ cations configure into a regular Ag$_6$ octahedron, with Ag-Ag bonds constituting the edges. The NaS$_6$ octahedra are interconnected through edge- and face-sharing, creating a three-dimensional network. This structure can be viewed as the embedding of Ag$_6$ octahedra within the framework formed by edge-sharing NaS$_6$ octahedra. At each face of the Ag$_6$ octahedra, an S$^{2-}$ anion bonds equally to three Ag$^{+}$ cations. The bond length between the S$^{2-}$ anion and the Ag$^{+}$ cation is measured at 2.4601 \AA\ for NaAg$_3$S$_2$ and 2.4768 \AA\ for KAg$_3$S$_2$, closely aligning with the Ag-S bond length of 2.3881 \AA\ calculated using PBEsol for $\alpha$-Ag$_2$S, indicating a robust bonding interaction between Ag$^{+}$ and S$^{2-}$. In contrast, the Ag-Ag bond lengths in NaAg$_3$S$_2$ (3.0654 \AA) and KAg$_3$S$_2$ (3.0851 \AA) are significantly greater than those found in Ag$_3$Ge$_5$P$_6$ (2.8426 \AA) and the bulk silver (2.88 \AA)~\cite{doi:10.1021/ic00339a009}. This elongation is attributed to the presence of the bridging S$^{2-}$ anion between the Ag-Ag bonds~\cite{doi:10.1021/ic50186a032}.

    The crystal orbital Hamilton population (COHP) and integrated COHP (iCOHP) of NaAg$_3$S$_2$ and Ag$_3$Ge$_5$P$_6$ are presented in Figure~\ref{crystalstructure}(b). The Ag-Ag bond of NaAg$_3$S$_2$ exhibits stronger antibonding states than that of Ag$_3$Ge$_5$P$_6$ in the energy range from -5 eV to the Fermi level. Therefore, the -iCOHP of NaAg$_3$S$_2$ is much smaller than that of Ag$_3$Ge$_5$P$_6$. The Ag-Ag bond lengths and the -iCOHP of Ag$_3$Ge$_5$P$_6$, NaAg$_3$S$_2$, and KAg$_3$S$_2$ are depicted in Figure~\ref{2ndFC} and summarized in Table~\textcolor{red}{S1}. Notably, the Ag-Ag bond lengths in all NaAg$_3$S$_2$-type compounds are indeed longer than that in Ag$_3$Ge$_5$P$_6$, and the -iCOHP values for the Ag-Ag bonds in NaAg$_3$S$_2$-type compounds are substantially lower than that in Ag$_3$Ge$_5$P$_6$, indicating significantly weaker Ag-Ag bonds in the former two compounds.

	\begin{figure*}[tph!]
	\includegraphics[width=1.0\linewidth]{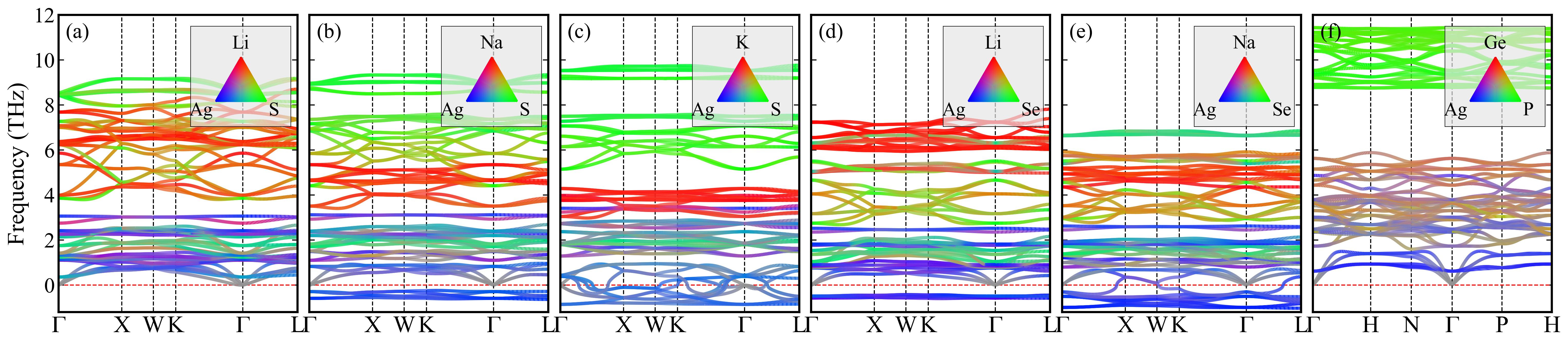}
	\caption{(a)-(f) are the phonon dispersion of LiAg$_3$S$_2$, NaAg$_3$S$_2$, KAg$_3$S$_2$, LiAg$_3$Se$_2$, NaAg$_3$Se$_2$, and Ag$_3$Ge$_5$P$_6$ at 0 K, respectively. Red, blue, and green colors represent the contributions of $A$, Ag, and $X$ (P) atoms to the phonon branches, respectively.}
	\label{phonon-0k}
    \end{figure*}

	\begin{figure*}[tph!]
	\includegraphics[width=1.0\linewidth]{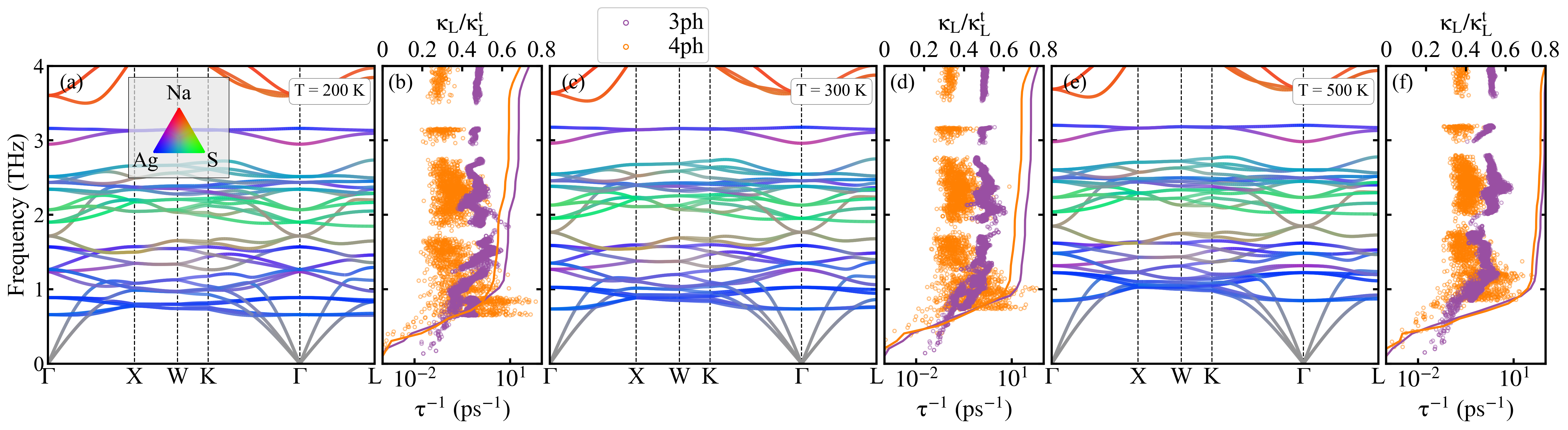}
	\caption{The phonon dispersion, 3ph- and 4ph-scattering, and the relative cumulative $\kappa_{\rm L}$ (the ratio of the cumulative $\kappa_{\rm L}$ to total $\kappa_{\rm L}$) of NaAg$_3$S$_2$ at 200, 300, and 500 K, respectively. Red, blue, and green colors represent the atomic contributions of Na, Ag, and S to each phonon band, respectively.}
	\label{phonon}
    \end{figure*}    
    	
	We conducted a further search for thermodynamically stable NaAg$_3$S$_2$-type compounds within the family of $A$Ag$_3$$X_2$ ($A$ = Li, Na, K, Rb, Cs and $X$ = S or Se) compounds through prototype structure decoration. Notably, the compounds LiAg$_3$S$_2$, LiAg$_3$Se$_2$, and NaAg$_3$Se$_2$ have not yet been experimentally reported, while KAg$_3$Se$_2$, RbAg$_3$S$_2$, RbAg$_3$Se$_2$, CsAg$_3$S$_2$, and CsAg$_3$Se$_2$ crystallize in the CsAg$_3$S$_2$-type structure (space group $C2/m$, No. 12). Thus, we first compared the total energies of the NaAg$_3$S$_2$-type and CsAg$_3$S$_2$-type structures for all $A$Ag$_3$$X_2$ compounds. As illustrated in Figure~\textcolor{red}{S1} of the supporting information, LiAg$_3$S$_2$, NaAg$_3$S$_2$, and NaAg$_3$Se$_2$ exhibit a preference for the NaAg$_3$S$_2$-type structure, while the remaining compounds favor the CsAg$_3$S$_2$-type structure. Moreover, the energy difference between these two structures increases linearly with the ionic radius of $A$, suggesting that compounds with larger $A$ ions are more inclined to adopt the CsAg$_3$S$_2$-type structure over the NaAg$_3$S$_2$-type structure. This implies that the experimentally observed $Fd\bar{3}m$ phase of KAg$_3$S$_2$ is a metastable phase. Convex hull calculations indicate that LiAg$_3$Se$_2$, NaAg$_3$S$_2$, and NaAg$_3$Se$_2$ are thermodynamically stable (i.e., they lie on the convex hull), whereas LiAg$_3$S$_2$ and KAg$_3$S$_2$ are metastable (i.e., positioned above the convex hull with a convex hull distance of less than 50 meV/atom) at 0 K, when assessed against their competing phases, as determined by the OQMD~\cite{OQMD1,OQMD2}, as shown in Table~\textcolor{red}{S1}. Similar to NaAg$_3$S$_2$ and KAg$_3$S$_2$, the Ag-Ag bond lengths and -iCOHP for all other NaAg$_3$S$_2$-type compounds are significantly smaller than those of Ag$_3$Ge$_5$P$_6$, as depicted in Figure~\ref{2ndFC} and Table~\textcolor{red}{S1} of the supporting information. Additionally, we compared the second-order interatomic force constants (2ndIFC) for the Ag-Ag bond, which are directly correlated to vibrational frequencies. To quantify the magnitude of these force constants, we utilized the root mean square (RMS) of the elements of the 2ndIFC tensor ($\Phi_{ij}$)~\cite{qin2018accelerating}. It is evident that all NaAg$_3$S$_2$-type compounds exhibit a significantly smaller RMS value of $\Phi_{ij}$ for the Ag-Ag bond compared to that of Ag$_3$Ge$_5$P$_6$.

	\begin{figure}[tph!]
	\includegraphics[width=1.0\linewidth]{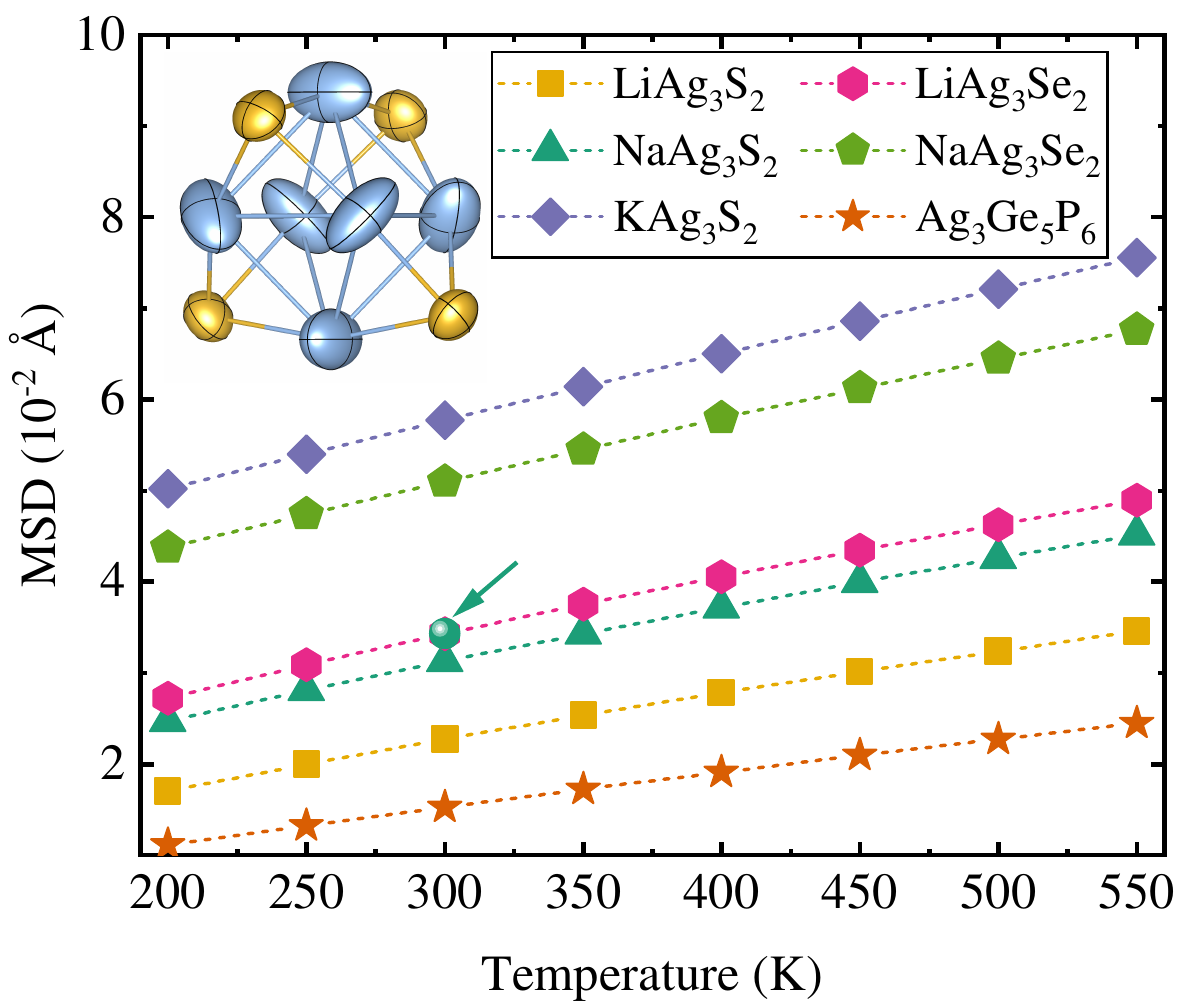}
	\caption{The calculated MSD of atom as a function of temperature. The green circle highlighted by a green arrow shows the experimental value of NaAg$_3$S$_2$. The inset depicts the anisotropic thermal displacement ellipsoids of Ag and S atoms at 300 K.}
	\label{msd}
    \end{figure}

	\vspace{0.5 cm}
	\noindent \textbf{Phonon dispersion at 0 K and 300 K.} The phonon dispersion and $\kappa_{\rm L}$ of LiAg$_3$S$_2$, LiAg$_3$Se$_2$, NaAg$_3$S$_2$, NaAg$_3$Se$_2$, KAg$_3$S$_2$, and Ag$_3$Ge$_5$P$_6$ were computed using a state-of-the-art calculation method~\cite{doi:10.1080/14786440408520575,PhysRevLett.17.753,werthamer1970self}, which accounts for three- and four-phonon interactions. This computational approach elucidates the effects of Ag-Ag bonding on phonon dispersion and $\kappa_{\rm L}$ by comparing five NaAg$_3$S$_2$-type compounds with Ag$_3$Ge$_5$P$_6$. The phonon dispersion at 0 K for the NaAg$_3$S$_2$-type compounds and Ag$_3$Ge$_5$P$_6$, calculated using the 2ndFC, is presented in Figure~\ref{phonon-0k}. All of these compounds exhibit low-frequency and flat-band phonons primarily associated with Ag atoms. Flat-band phonons indicate highly localized atomic vibrations and are characteristic of rattling phonon modes, which are typically observed in compounds that contain weakly bonded single or multiple atoms~\cite{xia2019impact,voneshen2013suppression,RN19,Jiangang2016,https://doi.org/10.1002/anie.201605015}. The atomic vibrations corresponding to these flat-band phonon modes at the $\Gamma$ point of the Brillouin zone are illustrated in Figure~\textcolor{red}{S2} of the supporting information. It is noteworthy that the vibrations of the atoms in both NaAg$_3$S$_2$ and Ag$_3$Ge$_5$P$_6$ are predominantly confined to the Ag atoms. In the Ag$_6$ octahedron, only three out of six Ag atoms participate in rotation, forming a head-to-tail triangle that is stretching in coordination with the remaining three Ag atoms. Consequently, the stiffness of this stretching mode, which determines the frequency of the phonon mode, is reliant on the strength of the Ag-Ag bond. Specifically, a weaker Ag-Ag bond strength results in a lower phonon frequency and a larger vibrational amplitude. The fourth and fifth phonon branches vibrate in a similar pattern, but in the opposite direction. These vibrational characteristics suggest that these Ag atoms behave as rattling phonon modes, akin to the behavior of a single atom within the cage-like structures. In addition, the low-lying and flat-band phonon modes of Ag$_3$Ge$_5$P$_6$ closely resemble the rattling modes observed in NaAg$_3$S$_2$.

    Furthermore, as depicted in Figure~\textcolor{red}{S3}, Debye-Einstein fitting analysis of the low-temperature specific heat capacity C$_p$ revealed the presence of two Einstein modes: $\Theta_{E_1}$ = 30 K ($\sim$ 0.63 THz) and $\Theta_{E_2}$ = 65.4 K ($\sim$ 1.35 THz). Notably, the low-energy mode $\Theta_{E_1}$ aligns well with the theoretically predicted low-lying optical phonons at approximately 0.73 THz (see Figure~\ref{phonon}(c)), thereby validating our computational results. This correlation explains why the flat-band phonon frequency of NaAg$_3$S$_2$ is lower than that of Ag$_3$Ge$_5$P$_6$. As the size of $A$ transitions from Li to K, the frequencies of the low-frequency flat-band phonons in the $A$Ag$_3X_2$ compounds decrease, and more flat-band phonons become unstable. This behavior is primarily attributed to the strain induced by the expansion of the $A$-$X$ bond length, rendering the NaAg$_3$S$_2$-type structure less favorable for larger $A$ cations, as discussed previously. The differences between $A$Ag$_3$S$_2$ and $A$Ag$_3$Se$_2$ can be traced to two factors: the variation in the anion radius and the bonding interaction difference between Ag and S and Ag and Se. These differences are rooted in the energy levels of the $p$-orbitals in the chalcogen anions, which influence the overlap between the Ag $d$ orbitals and the $p$ orbitals of the ions~\cite{woods2020wide,PhysRevB.29.1882}. Consequently, the $A$Ag$_3$Se$_2$ compounds exhibit smaller RMS values of $\Phi_{ij}$ compared to the $A$Ag$_3$S$_2$ compounds. Therefore, it is feasible to manipulate phonon behavior and heat transport by substituting the $A$ and $X$ ions.

	\begin{figure}[tph!]
		\includegraphics[width=1.0\linewidth]{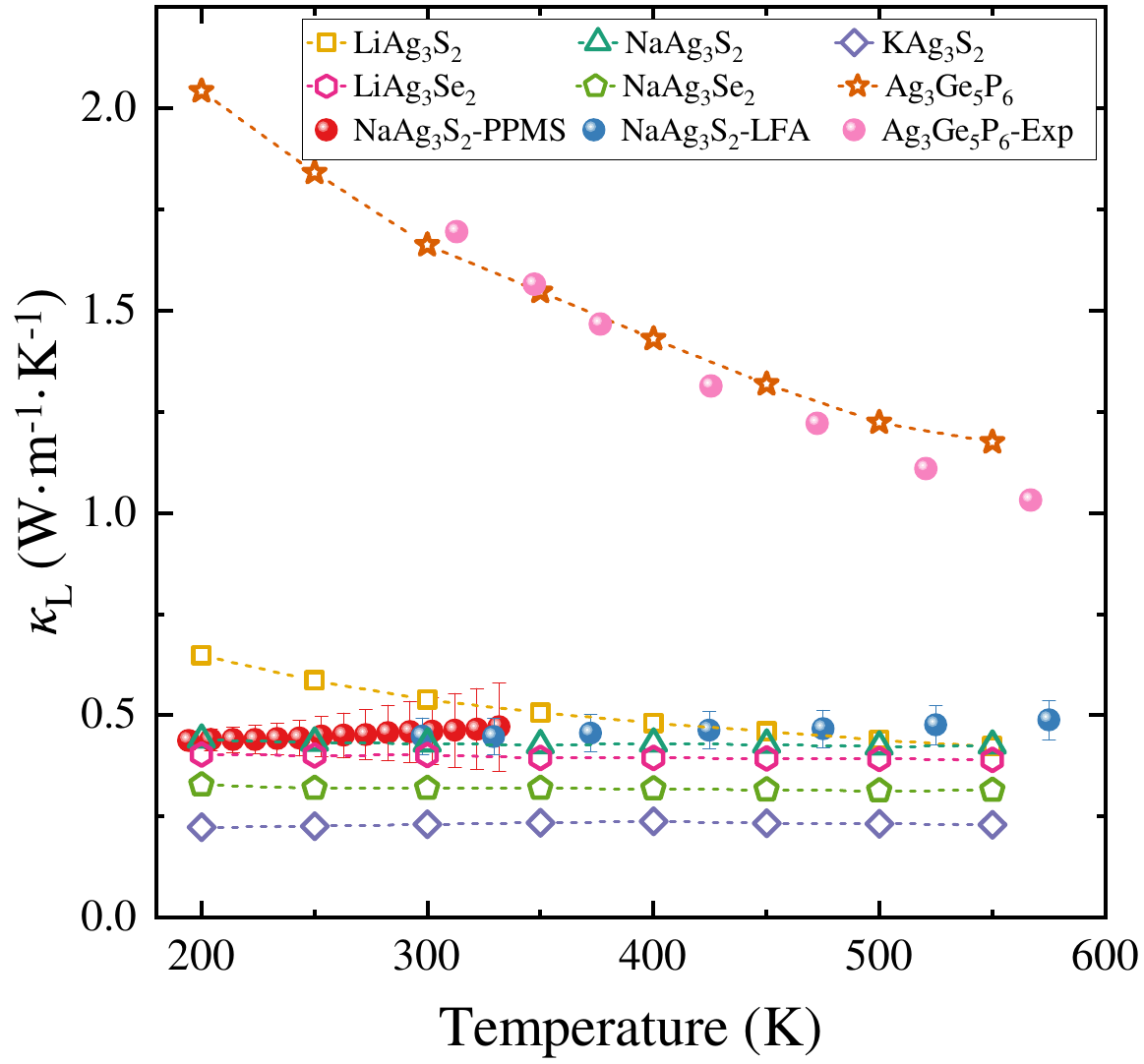}
		\caption{Experimental and computational $\kappa_{\rm L}$ of $A$Ag$_3X_2$ ($A$ = Li, Na, and K; $X$ = S and Se) and Ag$_3$Ge$_5$P$_6$ as a function of temperature. The experimental $\kappa_{\rm L}$ of Ag$_3$Ge$_5$P$_6$ is taken from the single crystal date in Ref.~\cite{doi:10.1021/acs.chemmater.7b02474}.}
		\label{kappa}
	\end{figure}

    The effect of temperature on phonon dispersion is investigated using self-consistent phonon (SCPH) theory~\cite{doi:10.1080/14786440408520575}, which incorporates the first-order corrections from fourth-order force constants~\cite{PhysRevLett.17.753,werthamer1970self}. The temperature-dependent phonon dispersion of NaAg$_3$S$_2$ across a range of 200 to 500 K is illustrated in Figure~\ref{phonon}. It is evident that temperature significantly influences the low-frequency phonon bands, particularly the flat-band phonons, while having negligible effects on the remaining phonon bands, especially the high-frequency ones. As the temperature increases, the frequencies of the flat-band phonons exhibit a rapid hardening, whereas their dispersion remains relatively constant, indicating a strong temperature dependence of the 2ndIFCs for Ag atoms. This phenomenon is further reflected in the MSD of the constituent atoms, which can be calculated from the renormalized second-order force constants to account for temperature effects. As shown in Figure~\ref{msd}, the magnitude of the Ag MSD increases linearly with rising temperature. Our calculated MSD values for NaAg$_3$S$_2$ align well with experimental observations. Additionally, all NaAg$_3$S$_2$-type compounds examined in this study exhibit larger Ag MSD values compared to Ag$_3$Ge$_5$P$_6$, consistent with our previous analysis that the NaAg$_3$S$_2$-type structure possesses significantly weaker Ag-Ag bonding.

\begin{figure}[tph!]
		\includegraphics[width=1.0\linewidth]{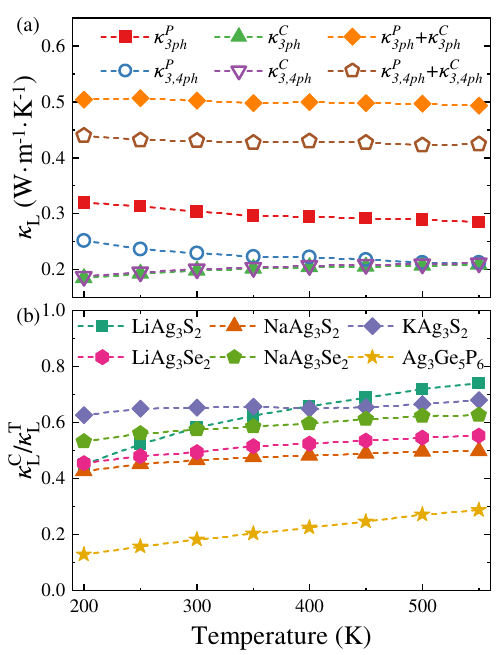}
		\caption{(a) The temperature-dependent $\kappa_{\rm L}$ with the contribution from phonon populations and coherences in the framework of three- and four-phonon scattering. (b) The ratio of $\kappa_{\rm L}$ from coherent contribution ($\kappa_{\rm L}^{\rm C}$) to total $\kappa_{\rm L}$ ($\kappa_{\rm L}^{\rm T}$) for all the studied compounds.}
		\label{kappa_all}
	\end{figure}

	\vspace{0.5 cm}
	\noindent \textbf{Lattice thermal conductivity.} Figure~\ref{kappa} presents the calculated $\kappa_{\rm L}$ for NaAg$_3$S$_2$-type compounds alongside that of Ag$_3$Ge$_5$P$_6$, employing a consistent computational methodology, see Method section. The $\kappa_{\rm L}$ value for Ag$_3$Ge$_5$P$_6$ obtained from our calculations aligns well with experimental measurements, exhibiting a weak dependence on temperature, characterized by T$^{-0.59}$. Notably, all NaAg$_3$S$_2$-type compounds display considerably lower $\kappa_{\rm L}$ values and a weaker temperature dependence compared to Ag$_3$Ge$_5$P$_6$. Furthermore, the $\kappa_{\rm L}$ of the NaAg$_3$S$_2$-type compounds decreases with increasing atomic number of both $A$ and $X$ as one moves from S to Se. Among these compounds, KAg$_3$S$_2$ exhibits the lowest $\kappa_{\rm L}$ at 0.23 Wm$^{-1}$K$^{-1}$ over the temperature range of 200 to 550 K, while LiAg$_3$S$_2$ displays the highest $\kappa_{\rm L}$ value at 0.54 Wm$^{-1}$K$^{-1}$ at 300 K. It is important to note that our calculated $\kappa_{\rm L}$ encompasses contributions from both particle-like heat transport (the diagonal component of the heat flux operator, denoted as $\kappa_{\rm L}^{P}$) and wave-like interband tunneling (the off-diagonal component of the heat flux operator, designated as $\kappa_{\rm L}^{C}$)~\cite{simoncelli2019unified}. Our findings indicate that the off-diagonal component makes a relatively larger contribution to the total $\kappa_{\rm L}$ in the NaAg$_3$S$_2$-type compounds, as depicted in Figure~\ref{kappa_all}. This observation is consistent with the enhanced anharmonicity associated with the weaker Ag-Ag and Ag-S bonds present in these compounds.

    The notably low $\kappa_{\rm L}$ can be attributed to strong three-phonon and four-phonon scattering processes, as illustrated in Figure~\ref{phonon}. The relative contribution of accumulated $\kappa_{\rm L}$ (expressed as $\kappa_{\rm L}/{\kappa}_{\rm L}^{\rm t}$, where $\kappa_{\rm L}^{\rm t}$ denotes the total $\kappa_{\rm L}$ reveals that phonons with frequencies below 2 THz account for 60\% of the total $\kappa_{\rm L}$, with the most significant contributions originating from the flat-band phonons at approximately 1 THz, where both three- and four-phonon scattering processes play crucial roles. In this low-frequency region ($\sim$ 1 THz), the contribution from four-phonon scattering becomes significant, coinciding with a notable overlap of acoustic and optical modes. In this context, the intensity of four-phonon scattering may exceed that of three-phonon scattering, leading to a pronounced reduction in $\kappa_{\rm L}$. These flat branches near 1 THz possess smaller group velocities and larger scattering phase spaces, contributing to the observed decrease in $\kappa_{\rm L}$. As the temperature rises, the frequency of the flat-band phonons increases, and the four-phonon scattering rate experiences a slight decline, resulting in a gradual decrease in $\kappa_{\rm L}^{P}$, as shown in Figure~\ref{kappa_all}.

    To verify our predictions, we experimentally synthesized NaAg$_3$S$_2$ polycrystals and measured its $\kappa_{\rm L}$ from 200 to 550 K using both PPMS and LFA methods, with additional experimental details provided in the Methods section. As illustrated in Figure~\ref{kappa}, the $\kappa_{\rm L}$ of NaAg$_3$S$_2$ exhibits nearly temperature-independent behavior over the range of 200 to 550 K, maintaining a value of approximately 0.45 Wm$^{-1}$K$^{-1}$, indicative of glass-like behavior. Notably, our calculated $\kappa_{\rm L}$ for NaAg$_3$S$_2$ aligns very closely with the experimental measurements, not only in terms of the absolute values at specific temperatures but also concerning the temperature dependence observed from 200 to 550 K. Note the calculation of $\kappa_{\rm L}$ becomes unphysical below 200 K due to the emergence of imaginary frequencies in this temperature regime.

\begin{figure}[tph!]
		\includegraphics[width=1.0\linewidth]{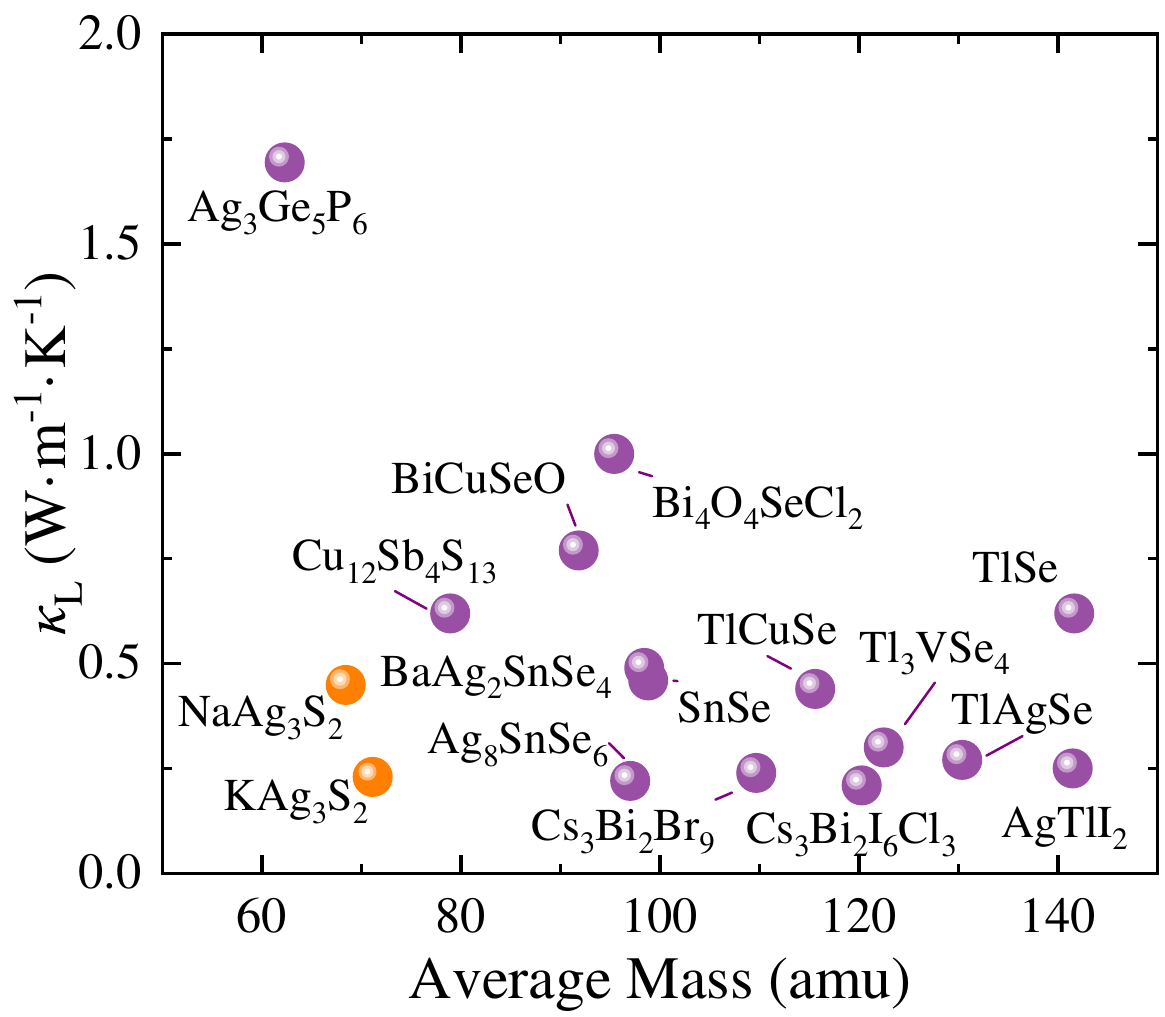}
		\caption{The dependence of $\kappa_{\rm L}$ on averaged mass ($\overline{M}$). The experimental $\kappa_{\rm L}$ values are taken from Ref.\cite{li2024phonon,mukhopadhyay2018two,li2016low,zeng2024pushing,acharyya2022glassy,lu2013high,doi:10.1021/acs.chemmater.7b02474,pathakdeciphering,zhu2017high,gibson2021low,zhou2021polycrystalline,dutta2019ultralow,li2019ultralow,lin2021ultralow} and where NaAg$_3$S$_2$ and KAg$_3$S$_2$ data were obtained from experimental and computational results in this paper, respectively.}
		\label{kappacompare}
	\end{figure}

	\vspace{0.5 cm}
	\noindent \textbf{Unique attributes of NaAg$_3$S$_2$-type compounds.}
    As discussed previously, low group velocities ($\nu_g$) and short lifetimes ($\tau$) typically result in reduced $\kappa_{\rm L}$. Consequently, low $\kappa_{\rm L}$ values are predominantly observed in compounds containing heavy elements~\cite{li2024phonon,mukhopadhyay2018two,li2016low,zeng2024pushing}, such as Tl, Pb, and Bi. These heavier elements not only possess higher atomic masses but also have lower electronegativities, which enhances their propensity to form weaker chemical bonds. However, NaAg$_3$S$_2$-type compounds exhibit a considerably lower average mass ($\overline{M}$) compared to other low-$\kappa_{\rm L}$ compounds, as illustrated in Figure~\ref{kappacompare}. Despite this, NaAg$_3$S$_2$ and KAg$_3$S$_2$ demonstrate significantly lower $\kappa_{\rm L}$ values than other copper- and silver-containing compounds with similar $\overline{M}$, making them competitive with compounds featuring heavier $\overline{M}$. It is noteworthy that Cu- and Ag-containing compounds generally also exhibit low $\kappa_{\rm L}$ values due to the presence of fully filled antibonding orbitals~\cite{https://doi.org/10.1002/adfm.202108532}. As shown in Table~\textcolor{red}{S2} of the supporting information, NaAg$_3$S$_2$-type compounds possess relatively low $\nu_g$ ($\sim$ 1500 m/s), comparable to those of heavy $\overline{M}$ compounds such as Ag$_8$SnSe$_6$ and Ag$_2$Te~\cite{li2016low}, despite their lower $\overline{M}$. This can be attributed to the weaker Ag-S and Ag-Ag chemical bonds present in these compounds. Another consequence of these weak bonds is the strong anharmonicity, characterized by elevated phonon-phonon scattering rates (1/$\tau$), which includes both three-phonon and four-phonon scattering processes. Notably, the four-phonon scattering rate in NaAg$_3$S$_2$ within the low-frequency region is significantly larger than that of the three-phonon scattering, which is typically dominant in phonon-phonon interactions. Particularly, the four-phonon scattering rates of the flat-phonon bands in NaAg$_3$S$_2$ reach values as high as 10 ps\(^{-1}\), indicating strong anharmonicity that exceeds their vibrational frequencies and thus breaches the Ioffe-Regel limit. This leads to a substantial contribution to $\kappa_{\rm L}$ from the off-diagonal component, as illustrated in Figure~\ref{kappa_all}(b).

	\section{CONCLUSIONS}
    In this work, we proposed and demonstrated an effective strategy for weakening Ag-Ag chemical bonds within a Ag$_6$ octahedron by enhancing the antibonding states below the Fermi level through the introducing an anion at the bridging position. Our material search based on this local coordination environment and new materials discovering based on structural decoration has yielded two existing and three synthesizable NaAg$_3$S$_2$-type compounds. The combination of low sound velocity and strong anharmonicity, induced by the weak Ag-S and Ag-Ag chemical bonds, significantly reduces the $\kappa_{\rm L}$ of NaAg$_3$S$_2$-type compounds. The hardening of the flat-band phonon modes associated with Ag atoms, along with the increased contribution from phonon tuning at elevated temperatures, results in a nearly temperature-independent $\kappa_{\rm L}$ across the temperature range of 200 to 550 K. Our experimental validation confirms that the $\kappa_{\rm L}$ of NaAg$_3$S$_2$ is approximately 0.45 Wm$^{-1}$K$^{-1}$ within the temperature range of 200 to 550 K. This finding underscores the effectiveness of our proposed material design strategy, which, founded on principles of chemical bonding, holds promise for application in other systems. The insights gained from this study not only advance our understanding of thermal transport mechanisms in low- and glass-like $\kappa_{\rm L}$ materials but also pave the way for the discovery of new compounds exhibiting desired thermal transport properties.

	\hspace{0.5cm}

		\noindent $\blacksquare$ \textbf{Supporting information} \\
		
		\noindent The Supporting Information is available free of charge on the ACS Publications website at DOI: \\

		\hspace{0.2cm}

		\noindent \textbf{Notes} \\
		The authors declare no competing financial interest.\\
		
		\hspace{0.5cm}
		
		\section{ACKNOWLEDGMENTS}
        X.S. acknowledges funding from the European Union's Horizon 2020 research and innovation program under the Marie Sklodowska-Curie grant agreement No. 101034329 and the WINNINGNormandy Program supported by the Normandy Region.
		Z.X. and J.H. acknowledge the support received from the Fundamental Research Funds for the Central Universities of China (USTB) and the National Natural Science Foundation of China (Grant No. 12304115). They also appreciate the computing resources provided by the USTB MatCom at the Beijing Advanced Innovation Center for Materials Genome Engineering. Y. X. acknowledges (1) the support from the US National Science Foundation through award 2317008, (2) the support from the Faculty Development Program at Portland State University, and (3) the computing resources provided by Bridges2 at Pittsburgh Supercomputing Center (PSC) through allocations mat220006p and mat220008p from the Advanced Cyber-infrastructure Coordination Ecosystem: Services \& Support (ACCESS) program, which is supported by National Science Foundation grants \#2138259, \#2138286, \#2138307, \#2137603, and \#2138296.

\bibliography{ref}


\end{document}